\documentclass[12pt,showpacs,showkeys,amsmath,amssymb]{revtex4}
\usepackage{amsmath,amsfonts,amsthm,amscd,amssymb,latexsym}
\usepackage{bm}
\usepackage{dcolumn}
\usepackage{graphicx}
\usepackage{epstopdf}
\usepackage{color}
\usepackage{epsf}
\usepackage{epsfig}
\usepackage{graphicx, epic, eepic, color}

\newcommand{\beq}{\begin{equation}}
\newcommand{\eeq}{\end{equation}}

\begin{document}

\title{Gravitomagnetic Stern--Gerlach Force}

\author{Bahram \surname{Mashhoon}$^{1,2}$}
\email{mashhoonb@missouri.edu}

\affiliation{$^1$Department of Physics and Astronomy, University of Missouri, Columbia, Missouri 65211, USA\\
$^2$School of Astronomy, Institute for Research in Fundamental
Sciences (IPM), P. O. Box 19395-5531, Tehran, Iran\\
}

\date{\today}

\begin{abstract}
A heuristic description of the spin-rotation-gravity coupling is presented and the implications of the corresponding gravitomagnetic Stern--Gerlach force are briefly mentioned. It is shown, within the framework of linearized general relativity, that the gravitomagnetic Stern--Gerlach force reduces in the appropriate correspondence limit to the classical Mathisson spin-curvature force. 
\end{abstract}

\pacs{03.30+p, 04.20.Cv}
\keywords{spin-vorticity coupling, spin-gravity coupling}

\maketitle

\section{Introduction}

Consider a free test particle of mass $m$ moving with velocity $\mathbf{V}$ in an inertial frame of reference in Minkowski spacetime. The free particle moves on a straight line with constant velocity forever. Here, the Minkowski metric is 
\begin{equation}\label{I1} 
-ds^2 = \eta_{\mu \nu}\,dX^\mu dX^\nu\,, \qquad  X^\mu = (ct, X, Y, Z)\,,
\end{equation}
where Greek indices run from 0 to 3, while Latin indices run from 1 to 3. The Minkowski metric tensor $\eta_{\mu \nu}$ is given by diag$(-1, 1, 1, 1)$. Throughout this paper, we use the convention that c = 1, unless specified otherwise. The equation of motion of the particle is obtained via the variational principle of stationary action $\delta \mathcal{S} = 0$, where 
\begin{equation}\label{I2} 
\mathcal{S} = \int -m\,ds = \int\mathcal{L}\,dt\,, \qquad \mathcal{L} = -m (1-V^2)^{1/2}\,.
\end{equation}
The corresponding Hamiltonian is $\mathcal{H}_0 = \gamma \,mc^2$, where $\gamma$ is the Lorentz factor. 

Let us now imagine that the static inertial observer at the origin of the spatial coordinates in Minkowski spacetime decides to refer the motion of the free particle to axes that rotate with angular velocity $\Omega (t)$ about the $Z$ axis. This static observer thus becomes noninertial and its new reference frame has coordinates $(ct, \mathbf{r})$, where $\mathbf{r} = (x, y, z)$.  Then, $\mathbf{V} = \mathbf{v} + \boldsymbol{\Omega}(t) \times \mathbf{r}$, where $\mathbf{v} = d\mathbf{r}/dt$ is the velocity of the particle with respect to the new rotating axes. From
\begin{equation}\label{I3} 
\mathcal{L} = -m \left[1-(\mathbf{v} + \boldsymbol{\Omega} \times \mathbf{r})^2\right]^{1/2}\,,
\end{equation}
we find the canonical momentum 
\begin{equation}\label{I4} 
\mathbf{p} = \frac{\partial \mathcal{L}}{\partial \mathbf{v}}  = \gamma m (\mathbf{v} + \boldsymbol{\Omega} \times \mathbf{r})\,
\end{equation}
and the Hamiltonian~\cite{LL}
\begin{equation}\label{I5} 
\mathcal{H} = \mathcal{H}_0 - \boldsymbol{\Omega}(t) \cdot \mathbf{L}\,,
\end{equation}
where $\mathbf{L} = \mathbf{r} \times \mathbf{p}$ is the orbital angular momentum of the free point particle. 

If the particle carries with it an ``intrinsic" spin vector $\mathbf{S}$, then $\mathbf{S}$ remains constant along the straight trajectory of the particle in the inertial frame. However, with respect to the rotating coordinate system, $\mathbf{S}$ appears to precess with angular velocity $- \boldsymbol{\Omega}(t)$. Let $s_i$, $i = 1, 2, 3$, be the components of $\mathbf{S}$ with respect to the rotating axes; then, 
\begin{equation}\label{I6} 
\frac{ds_i}{dt} + \epsilon_{ijk} \Omega^j s^k = 0\,.
\end{equation}
On the other hand, for a true  intrinsic quantum spin vector with the commutation relations
\begin{equation}\label{I7} 
[s_p, s_q] = i \hbar \epsilon_{pqn} s^n\,
\end{equation}
that is invariant under the rotation of coordinates, the Heisenberg equation of motion for such a quantum observable,
\begin{equation}\label{I8} 
\hbar \frac{ds_k}{dt} = i [\mathcal{H}_{SR}, s_k]\,,
\end{equation}
results in Equation~\eqref{I6} if the spin-rotation Hamiltonian is of the form
\begin{equation}\label{I9} 
\mathcal{H}_{SR} = -\mathbf{S} \cdot \boldsymbol{\Omega}(t)\,.
\end{equation}
This is the Hamiltonian that accounts for the precessional motion of the spin in the quantum theory. 
It follows that in the quantum case there is an additional contribution to the classical Hamiltonian~\eqref{I5} such that the total Hamiltonian of the particle in the rotating frame is given by $\mathcal{H}+\mathcal{H}_{SR}$. Hence,    
\begin{equation}\label{I10} 
\mathcal{H}_{Total} = \mathcal{H}_0 - \boldsymbol{\Omega}(t) \cdot \mathcal{J}\,,
\end{equation}
where $\mathcal{J} := \mathbf{L} + \mathbf{S}$ is the total angular momentum of the free particle. This is a natural result, since $\mathcal{J}$ is the generator of rotations in the quantum theory. The energy eigenvalues as measured by the noninertial static observer include the spin-rotation coupling, which is a quantum inertial effect that is independent of mass of the particle. In the classical limit, $\hbar \to 0$ and we recover Equation~\eqref{I5}.

Spin-rotation coupling is a general phenomenon that is due to the inertia of intrinsic spin. Physical states in quantum theory are described by mass and spin, which characterize the irreducible unitary representations of the inhomogeneous Lorentz group. The inertial properties of mass are well known. Phenomena associated with the spin-rotation coupling  reveal the inertial properties of intrinsic spin.
Spin-rotation coupling has extensive observational support~\cite{BMB, BM3, BM5, Mash, HN, BM5a, SoTi, LR, PJA, MNHS, BM4, Ash,HaMa, AnMa, BMash, Bliokh:2015yhi, SP, Pa, ShHe, LaPa,  Pan:2011zza, Arminjon:2013kxa}. It has recently been observed directly in neutron interferometry~\cite{ Mashhoon:2005fe, Werner, HRW, DSH, MK, Mashhoon:2015nea, DDSH, DDKWLSH}. Furthermore, it has significant applications in spintronics~\cite{MISM, MISM2,  MIM, CB, PAP, IMM,  NaTa, Hama, MIHSM,Taka, KYM, OMOS, MOM, KMN, KMMN, Kaz}. For further discussion and references, see~\cite{Mashhoon:1997qc, Mashhoon:2003ax}.

\subsection{Spin-Vorticity Coupling}

Consider a laboratory experiment involving a rotating system, which creates a congruence  in spacetime. As a body rotates, we expect that the intrinsic spins of the constituent particles all remain fixed with respect to the local inertial frame; therefore, the intrinsic spins all appear to precess with respect to the body-fixed frame. In the continuum limit, it may be that the local angular velocity of motion becomes dependent on position, in which case the spin-rotation coupling naturally goes over to the spin-vorticity coupling~\cite{MOM, OMOS}
\begin{equation}\label{I11} 
\mathcal{H}_{SV} = - \frac{1}{2} \,\mathbf{S} \cdot \boldsymbol{\omega}\,,
\end{equation}
where $\boldsymbol{\omega}$ is the vorticity  
\begin{equation}\label{I12} 
\boldsymbol{\omega} = \nabla \times \mathbf{V}\,
\end{equation}
and $\mathbf{V}$ is the velocity field of the congruence. If the angular velocity is spatially uniform such that $\mathbf{V} = \boldsymbol{\Omega} \times \mathbf{r}$, then $\boldsymbol{\omega} = 2 \,\boldsymbol{\Omega}$ and $\mathcal{H}_{SV}$ reduces to $\mathcal{H}_{SR}$. For a description of moving macroscopic matter in continuum mechanics, see section E.4.1 of Ref.~\cite{HeOb}. For recent work on spin-vorticity coupling, see~\cite{KMN, KMMN}.

\subsection{Stern--Gerlach Force due to Spin-Vorticity Coupling}

In general, vorticity depends on position and we might then expect the appearance of an attendant Stern--Gerlach force as well; that is, 
\begin{equation}\label{I13} 
f_\mu = -\partial_\mu (\mathcal{H}_{SV}) =  \frac{1}{2} \,\mathbf{S} \cdot \nabla_{\mu}\,\boldsymbol{\omega}\,.
\end{equation}
Such a spin-dependent force could then lead to the generation of a spin current. This idea was apparently first proposed in Ref.~\cite{MIHSM} and received experimental confirmation in~\cite{Taka, KYM, OMOS}.   For the extension of these ideas to fluid spintronics, see~\cite{MOM} and the references cited therein.  Moreover, Ref.~\cite{Kaz} deals with the application of spin-vorticity coupling in fluid dynamics.

\section{Spin-Gravity Coupling}

Within the framework of linearized general relativity, we use here the approximation scheme known as gravitoelectromagnetism (GEM) that is based on the well-known analogy with Maxwell's electrodynamics. We are interested in the weak exterior field of a compact rotating astronomical source with mass $M$ and proper angular momentum $\mathbf{J}$. The spacetime metric, $-ds^2 = g_{\mu \nu} \,dx^\mu dx^\nu$,  is given in a Cartesian system of coordinates $x^\alpha = (ct, \mathbf{x})$ by~\cite{Mashhoon:2003ax}
\begin{equation}\label{G1} 
-ds^2=-c^2\left(1-2\frac{\Phi}{c^2}\right)dt^2-\frac{4}{c}({\bf 
A}\cdot d{\bf x})dt+\left(1+2\frac{\Phi}{c^2}\right) \delta_{ij}dx^idx^j\,,
\end{equation}
which represents Minkowski spacetime plus a linear perturbation due to the source. That is, $g_{\mu \nu} = \eta_{\mu \nu} + h_{\mu \nu}$. We neglect all metric perturbation terms of $O(c^{-4})$ in this weak-field and slow-motion approximation method. In Equation~\eqref{G1}, $\Phi(t, \mathbf{x})$ is the gravitoelectric potential and $\mathbf{A}(t, \mathbf{x})$ is the gravitomagnetic vector potential. For the exterior field of a rotating astronomical mass, for instance, $-\Phi$ is the Newtonian gravitational potential and $\mathbf{A}$ is due to mass current and vanishes in the Newtonian limit ($c \to \infty$). Very far from the rotating source, 
\begin{equation}\label{G2}
 \Phi \sim \frac{GM}{r}\,,\qquad \mathbf{A}\sim 
\frac{G}{c}\frac{\mathbf{J}\times \mathbf{x}}{r^3}\,,
\end{equation}
where $r=|\mathbf{x}|$. The GEM potentials satisfy the transverse gauge condition 
\begin{equation}\label{G3}
 \frac{1}{c}\frac{\partial \Phi}{\partial 
t}+\nabla \cdot \left(\frac{1}{2}\mathbf{A}\right)=0\,.
\end{equation}
Moreover, in analogy with electrodynamics, the GEM fields are defined by
\begin{equation}\label{G4} 
\mathbf{E}= -\nabla \Phi -\frac{1}{c}\frac{\partial}{\partial t}\left(\frac{1}{2}\mathbf{A}\right)\,,\qquad \mathbf{B} =\nabla \times \mathbf{A}\,,
\end{equation}
in terms of which Einstein's field equations in this case become formally similar to Maxwell's equations~\cite{AE}. For discussions of the non-Newtonian gravitomagnetic effects, see~\cite{Mashhoon:2003ax, RuTa}. 

We are interested in the motion of a classical spinning point particle in the GEM field. The relevant equations in this case are the Mathisson--Papapetrou (``pole-dipole") equations~\cite{Math, Papa}, namely, 
\begin{equation}\label{G5} 
\frac{DP^\mu}{d\tau} = F^\mu\,, \qquad F^\mu = -\frac{1}{2} R^{\mu}{}_{\nu \alpha \beta}\,u^\nu S^{\alpha \beta}\,,
\end{equation}
\begin{equation}\label{G6} 
\frac{DS^{\mu \nu}}{d\tau} = P^\mu u^\nu - P^\nu u^\mu\,.
\end{equation}
In these equations, $F^\mu$, $F^\mu u_\mu = 0$,  is the Mathisson spin-curvature force~\cite{{Mashhoon:2008si}}, $u^\mu = dx^\mu/d\tau$ is the 4-velocity of the pole-dipole particle and $\tau$ is its proper time. Moreover, $P^\mu$ is the 4-momentum of the particle and $S^{\mu \nu}$ is its spin tensor that satisfies the Frenkel--Pirani supplementary condition~\cite{Frenkel, Pirani} 
\begin{equation}\label{G7} 
S^{\mu \nu}\,u_\nu = 0\,.
\end{equation}
It follows from these equations that
\begin{equation}\label{G8} 
P^\mu = m\,u^\mu + S^{\mu \nu}\,\frac{Du_\nu}{d\tau}\,,
\end{equation}
where $m := -P^\mu u_\mu$ is the mass of the spinning particle and is a constant of the motion. That is, differentiating $m = - P \cdot u$ and using $F\cdot u = 0$ together with Equation~\eqref{G8}, we find
\begin{equation}\label{G9} 
\frac{dm}{d\tau} =  - \left(m\,u^\mu + S^{\mu \nu}\,\frac{Du_\nu}{d\tau}\right) \frac{Du_\mu}{d\tau} = 0\,,
\end{equation}
since $u \cdot u = -1$ and $S^{\mu \nu}$ is antisymmetric. 
In the massless limit ($m \to 0$), the Mathisson--Papapetrou equations together with the Frenkel--Pirani supplementary condition behave properly; indeed, the spinning massless test  particle follows a null geodesic with spin parallel or antiparallel to its world line~\cite{Mashhoon}.    The Frenkel--Pirani supplementary condition is therefore appropriate for a point particle. 

Let us define the spin vector of the particle $S^\mu$ via
\begin{equation}\label{G10} 
S_\mu = -\frac{1}{2} \,\eta_{\mu \nu \rho \sigma}\,u^\nu S^{\rho \sigma}\,, \qquad S^{\alpha \beta} = \eta^{\alpha \beta \gamma \delta}\,u_\gamma S_\delta\,,
\end{equation}
where $\eta_{\alpha \beta \gamma \delta} = (-g)^{1/2} \epsilon_{\alpha \beta \gamma \delta}$ is the Levi-Civita tensor and  $\epsilon_{\alpha \beta \gamma \delta}$ is the alternating symbol with 
$\epsilon_{0123} = 1$. The Mathisson spin-curvature force now takes the form
\begin{equation}\label{G11} 
F_\mu =~ ^*R_{\mu \nu \rho \sigma}\,u^\nu \,S^{\rho}\,u^\sigma\,, \qquad ^*R_{\mu \nu \rho \sigma} = \frac{1}{2}\, \eta_{\mu \nu \alpha \beta}\,R^{\alpha \beta}{}_{\rho \sigma}\,,
\end{equation}
in terms of the dual Riemann tensor, and the spin dynamics is represented by
\begin{equation}\label{G12} 
(g_{\mu \nu} + u_\mu\, u_\nu)\,\frac{DS^\nu}{d\tau} = 0\,,
\end{equation}
so that $S^\mu$, $S^\mu\, u_\mu = 0$, is Fermi--Walker transported along the world line of the spinning particle~\cite{Pirani}. 

Consider now a pole-dipole particle held at rest in space in the exterior GEM field. Nongravitational torque-free forces are necessary to counter the Mathisson force as well as the attraction of gravity of the source in order to keep the particle fixed in space. The 4-velocity vector of the particle is then given by $u^\mu = (1 +\Phi/c^2) \delta^\mu_0$. A natural orthonormal tetrad frame $\lambda^{\mu}{}_{(\alpha)}$ adapted to the static test pole-dipole particle with $ u^\mu = \lambda^{\mu}{}_{(0)}$ is given in the $(ct, x, y, z)$ coordinate system by
\begin{equation}\label{G13} 
\lambda^{\mu}{}_{(0)} = (1+\Phi/c^2, 0, 0, 0)\,,
\end{equation}
\begin{equation}\label{G14} 
\lambda^{\mu}{}_{(1)} = (-2A_1/c^2, 1-\Phi/c^2, 0, 0)\,,
\end{equation}
\begin{equation}\label{G15} 
\lambda^{\mu}{}_{(2)} = (-2A_2/c^2, 0, 1-\Phi/c^2, 0)\,,
\end{equation}
\begin{equation}\label{G16} 
\lambda^{\mu}{}_{(3)} = (-2A_3/c^2, 0, 0,  1-\Phi/c^2)\,,
\end{equation}
where the tetrad axes are primarily along the local GEM coordinate axes. The projection of the spin vector on the adapted tetrad frame is given by
\begin{equation}\label{G17} 
S_{(\alpha)} = S_\mu\,\lambda^{\mu}{}_{(\alpha)}\,,
\end{equation}
which implies that $S_{(0)} = 0$ and
\begin{equation}\label{G18} 
\frac{dS_{(i)}}{d\tau} = \left[\frac{D\lambda^{\mu}{}_{(i)}}{d\tau}\,\lambda_{\mu (j)}\right]\,S^{(j)}\,.
\end{equation}
A straightforward calculation reveals that to linear order in the perturbation
\begin{equation}\label{G19} 
\frac{D\lambda^{\mu}{}_{(i)}}{d\tau}\,\lambda_{\mu (j)} = \partial_j\,A_i -  \partial_i\,A_j\,.
\end{equation}
Therefore, 
\begin{equation}\label{G20} 
\frac{dS_{(i)}}{d\tau} =  \epsilon_{ijk} B^j\,S^{(k)}\,
\end{equation}
and the spin vector precesses with an angular velocity given by the local gravitomagnetic field. We can regard the gravitomagnetic field in Equation~\eqref{G20} as the locally \emph{measured} field within our approximation scheme. That is, the GEM potentials can be combined into a 4-vector in analogy with electrodynamics and the corresponding GEM field tensor is then projected on the tetrad frame 
$\lambda^{\mu}{}_{(\alpha)}$ to obtain the measured gravitoelectric and gravitomagnetic fields at the location of the spinning particle. However, $\lambda^{\mu}{}_{(\alpha)}$ differs from $\delta^{\mu}_{\alpha}$ by terms that are linear in the spacetime perturbation; therefore, in our approximation method $\mathbf{E}$ and $\mathbf{B}$ are indeed the same as the measured fields. 

If the spin vector is of quantum origin and represents the intrinsic spin of the ``point" particle, then a spin-gravity Hamiltonian in terms of measured quantities is associated with its precession such that 
\begin{equation}\label{G21} 
\mathcal{H}_{SG} =  \frac{1}{c}\,\mathbf{S} \cdot \mathbf{B}\,.
\end{equation}
We assume here that a particle with intrinsic spin behaves in the correspondence limit like an ideal gyroscope. 
For instance, in connection with experiments in an Earth-based laboratory, to 
every Hamiltonian we must add the spin-rotation-gravity
contribution 
\begin{equation}\label{G22} 
\delta \mathcal{H} \approx -\boldsymbol{\Omega} _\oplus 
\cdot \mathbf{S}+ \boldsymbol{\Omega}_P\cdot \mathbf{S}\,,
\end{equation}
where
$\boldsymbol{\Omega }_\oplus$ and $\boldsymbol{\Omega}_P = \mathbf{B}_{\oplus}/c$ refer to the Earth's rotation frequency and the corresponding gravitomagnetic precession frequency, respectively. In fact, we have approximately
\begin{equation}\label{G23}
\boldsymbol{\Omega}_P=\frac{G}{c^2r^5}[3({\bf J}\cdot {\bf r})\,{\bf r}- {\bf J} r^2]\,.
\end{equation}
In the recent GP-B experiment~\cite{Francis1, Francis2},  the non-Newtonian gravitomagnetic field of the Earth has been directly measured and the prediction of general relativity has been verified at about the 19\% level. 

In Equation~\eqref{G22}, the difference in the energy of a spin-$1/2$ particle polarized vertically up and down relative to the surface of the Earth is given by $\hbar \Omega_\oplus \approx
10^{-19}\text{eV}$ and $\hbar \Omega_P\approx 10^{-29}\text{eV}$.  For recent attempts to measure the spin-gravity term, see~\cite{Tarallo:2014oaa, GPS}. Furthermore, the gravitomagnetic field depends upon position; therefore, there exists a gravitomagnetic Stern--Gerlach force $-\nabla (\boldsymbol{\Omega}_P\cdot \mathbf{S})$ on a spinning particle that is independent of its mass and hence violates the principle of equivalence and the universality of free fall. This force naturally leads to a differential deflection of polarized beams. For various implications of the spin-gravity coupling, see~\cite{BMc, BMd, Bah1, Bah2, Ior1, Bini:2004kz, CMP, Ran, Ior2, Mashhoon:2013jaa, Plyatsko:2017kux, Yuri, Obukhov:2017avp}.

We now wish to establish a general correspondence between the gravitomagnetic Stern--Gerlach force and the Mathisson spin-curvature force for a steady-state configuration. Projecting the gravitomagnetic Stern--Gerlach force, $f_\mu = -\partial_\mu \mathcal{H}_{SG}$ on the orthonormal tetrad frame $\lambda^{\mu}{}_{(\alpha)}$, we have
\begin{equation}\label{G24} 
f_{(0)} = 0\,, \qquad f_{(i)}  = -\frac{1}{c} \partial_i\,B_j \,S^{(j)}\,.
\end{equation}
Here, $f_{(0)}$ vanishes because the gravitomagnetic field is assumed to be independent of time. On the other hand, the Mathisson spin-curvature force projected on the tetrad frame $\lambda^{\mu}{}_{(\alpha)}$ is given by $F_{(0)} = 0$ and
\begin{equation}\label{G25} 
F_{(i)}  = c\,^*R_{(i)(0)(j)(0)} \,S^{(j)}\,.
\end{equation}
We want to show that $f_{(i)}$ reduces to $F_{(i)}$ in the correspondence limit. 

In an arbitrary gravitational field, one can project the Riemann tensor onto an orthonormal tetrad frame $\Theta^{\mu}{}_{\hat \alpha}$ adapted to an observer; the measured components of the curvature are then 
\begin{equation}\label{G26}
R_{\mu \nu \rho \sigma}\,\Theta^{\mu}{}_{\hat \alpha}\,
\Theta^{\nu}{}_{\hat \beta}\,\Theta^{\rho}{}_{\hat \gamma}\,\Theta^{\sigma}{}_{\hat \delta}\,.
\end{equation}
Taking the symmetries of the Riemann tensor into account, one can express Equation~\eqref{G26} in the standard manner as a $6\times 6$ matrix with indices that range over the set $\{01,02,03,23,31,12\}$. The general form of this matrix is 
\beq\label{G27}
\left[
\begin{array}{cc}
\mathbb {E} & \mathbb{H}\cr
\mathbb{H}^T & \mathbb{S}\cr 
\end{array}
\right]\,,
\eeq
where $\mathbb{E}$ and $\mathbb{S}$ are symmetric $3\times 3$ matrices and $\mathbb{H}$ is traceless. Here, $\mathbb{E}$, $\mathbb{H}$ and $\mathbb{S}$ represent the measured gravitoelectric,  gravitomagnetic and spatial components of the Riemann curvature tensor, respectively. If the spacetime is Ricci flat, then  Equation~\eqref{G27} takes the form
\beq\label{G28}
\left[
\begin{array}{cc}
\mathbb{E} & \mathbb{H}\cr
\mathbb{H} & -\mathbb{E}\cr 
\end{array}
\right]\,,
\eeq
where  $\mathbb{E}$ and $\mathbb{H}$ are now symmetric and traceless. That is, in the Ricci-flat case, the Riemann curvature tensor degenerates  into the Weyl conformal curvature tensor whose gravitoelectric and gravitomagnetic components are then 
\begin{equation}\label{G29} 
\mathbb{E}_{\hat a \hat b}= C_{\alpha\beta\gamma\delta}\,\Theta^\alpha{}_{\hat 0}\,\Theta^\beta{}_{\hat a}\, \Theta^\gamma{}_{\hat 0}\, \Theta^\delta{}_{\hat b}\,,\qquad \mathbb{H}_{\hat a \hat b}=C^*_{\alpha\beta\gamma\delta}\,\Theta^\alpha{}_{\hat 0}\,\Theta^\beta{}_{\hat a}\, \Theta^\gamma{}_{\hat 0}\, \Theta^\delta{}_{\hat b}\,,
\end{equation}
where  $C^*_{\alpha\beta\gamma\delta}$ is the unique dual of the Weyl tensor given by
\begin{equation}\label{G30} 
C^*_{\alpha\beta\gamma\delta}=\frac12\, \eta_{\alpha \beta}{}^{\mu\nu}\,C_{\mu\nu\gamma\delta}\,,
\end{equation}
since the right and left duals of the Weyl  tensor coincide.

In our GEM scheme, $g_{\mu \nu} = \eta_{\mu \nu} + h_{\mu \nu}$  and the gauge-invariant curvature tensor is given by 
\begin{equation}\label{G31}
R_{\mu \nu \rho \sigma} = \frac{1}{2} ( h_{\mu \sigma,\, \nu \rho} + h_{ \nu \rho, \,\mu \sigma} - h_{\nu \sigma,\, \mu \rho} - h_{\mu \rho, \,\nu \sigma})\,. 
\end{equation}
We recall that to lowest order $\lambda^{\mu}{}_{(\alpha)} \approx \delta^{\mu}_\alpha$ and hence in the exterior of a GEM source, we have the Weyl tensor in the form~\eqref{G28} with symmetric and traceless matrices given by
\begin{equation}\label{G32}
 \mathbb{E}_{ij} = - \frac{1}{c^2} \Phi_{,ij} + O(c^{-4}) = \frac{1}{c^2} E_{j,i} + O(c^{-4})\, 
\end{equation}
and
\begin{equation}\label{G33}
 \mathbb{H}_{ij} = - \frac{1}{c^2} \partial_i B_{j} + \frac{1}{c^3}\,\epsilon_{ijk}\,\frac{\partial E_k}{\partial t} + O(c^{-4})\,.
\end{equation}
It follows from these results and Equations~\eqref{G24}--\eqref{G25}  that for a \emph{stationary} GEM field, the gravitomagnetic Stern--Gerlach force in the correspondence limit is the same as the Mathisson spin-curvature force. 

The spin interactions discussed in this paper all involve Hamiltonians that are similar to that of the traditional Zeeman effect. Moreover, the gravitational Larmor theorem can be invoked to connect spin-gravity coupling with the spin-rotation coupling. The local equivalence between magnetism and rotational inertia was first established via Larmor's original theorem~\cite{Larmor}. The gravitational Larmor theorem is an expression of Einstein's local principle of equivalence of gravitation and inertia~\cite{Mashhoon:2003ax, Bahram}. 

Consider a steady-state configuration with exterior metric~\eqref{G1}. In this stationary gravitational field, the temporal coordinate can be subjected to a simple scale transformation of the form $t \mapsto (1+\Phi_0/c^2)t$, where $\Phi_0$ is a constant such that $|\Phi_0| \ll c^2$. The only consequence of this transformation is that $-g_{00}= 1-2\Phi_0 \mapsto1-2(\Phi-\Phi_0)/c^2$, while the other terms in the metric remain unchanged since we neglect all terms of $O(c^{-4})$. In a sufficiently small neighborhood around any event in   the exterior GEM spacetime, we can replace the metric by that of an accelerated system in Minkowski spacetime. The resulting metric is to linear order of the form $(\eta_{\mu\nu} +\ell_{\mu \nu})dX^\mu dX^\nu$, where~\cite{BMB, Mashhoon:2003ax} 
\begin{equation}\label{G34}
 \ell_{00} = -2\, \mathbf{a}_L \cdot \mathbf{X}\,, \qquad \ell_{0i} = (\boldsymbol{\Omega}_L \times \mathbf{X})_i\,, \qquad \ell_{ij} = 0\,.
\end{equation}
This has the form of a first-order perturbation where $\mathbf{a}_L$ is the constant translational acceleration and $\boldsymbol{\Omega}_L$ is the constant rotational frequency of the accelerated system. A comparison with the GEM metric reveals that the corresponding gravitoelectric and gravitomagnetic potentials are given by
\begin{equation}\label{G35}
 \Phi - \Phi_0 = - \mathbf{a}_L \cdot \mathbf{X}\,, \qquad \mathbf{A} = -\frac{1}{2}\,\boldsymbol{\Omega}_L \times \mathbf{X}\,.
\end{equation} 
We neglect the spatial curvature of the GEM metric in this analogy. Moreover, $\mathbf{E} = -\nabla \Phi = \mathbf{a}_L$ and $\mathbf{B} = \nabla \times \mathbf{A} = - \boldsymbol{\Omega}_L$ are the corresponding fields. It is clear that the spin-rotation Hamiltonian $\mathcal{H}_{SR} = -\mathbf{S} \cdot \boldsymbol{\Omega}_L$ corresponds to the spin-gravity Hamiltonian $\mathcal{H}_{SG} = \mathbf{S} \cdot \mathbf{B}$ via the gravitational Larmor theorem.

\section{Linear Gravitational Waves}

The general linear approximation of general relativity involves GEM fields of massive systems as well as linearized gravitational waves. The purpose of this section is to discuss spin-gravity coupling for linearized gravitational waves; in particular, we are interested in the corresponding Stern--Gerlach force. For related studies, see~\cite{Shen:2004uw, Ramos:2006sb, Hojman:2016mox, Papini:2019qxn, Papini:2020fgo, Ruggiero:2020nef, Ruggiero:2020oxo, Ruggiero:2021qnu} and the references cited therein.    

Consider a free linear gravitational radiation field, which can be expressed as a Fourier sum of plane monochromatic components each with frequency $\omega_g$ and wave vector $\mathbf{k}_g$, $\omega_g = c |\mathbf{k}_g|$. The gravitational potential of the radiation is given by the symmetric tensor $\bar{h}_{\mu \nu}$, which is a perturbation on the background Minkowski spacetime; that is, $g_{\mu \nu} = \eta_{\mu \nu} + \bar{h}_{\mu \nu}(x)$, where $x^\alpha = (ct, x, y, z)$. In the transverse-traceless (TT) gauge, $\bar{h}^{\mu \nu}{}_{,\nu} = 0$, $\bar{h}_{0\mu} = 0$ and $\bar{h}^{\mu}{}_{\mu} = 0$. In this gauge, the gravitational potentials $\bar{h}_{ij}(x)$ each satisfies the standard wave equation. 

For the sake of definiteness, let the incident radiation be a monochromatic plane wave propagating along the $x$ direction. Then, $\bar{h}_{ij}$ can be written as
\begin{equation}\label{W1}
(\bar{h}_{ij}) = 
\begin{bmatrix}
0 & 0 & 0 \\
0 & h_{+} & h_{\times} \\
0 & h_{\times} &- h_{+}
\end{bmatrix}
\,,
\end{equation}
where 
\begin{equation}\label{W2}
h_{+} = \tilde{h}_{+}\,\cos[\omega_g(t-x) +\varphi_{+}]\,, \qquad h_{\times} = \tilde{h}_{\times}\,\cos[\omega_g(t-x) +\varphi_{\times}]\,
\end{equation}
represent the $\oplus$ (``plus") and $\otimes$ (``cross") linear polarization states of the radiation. Here, $(\tilde{h}_{+}, \varphi_{+})$ and $(\tilde{h}_{\times}, \varphi_{\times})$ are constants associated with the independent states of the radiation field. 

It is a general result that in a spacetime with a metric of the form $-dt^2 + g_{ij}(x) dx^idx^j$, observers that remain permanently at rest in space follow geodesic world lines. Thus imagine this class of geodesic observers each at rest in space with a 4-velocity vector $e^{\mu}{}_{\hat 0} = \delta^\mu_0$ in the spacetime under consideration here. To each such observer, we associate an adapted orthonormal tetrad frame $e^{\mu}{}_{\hat \alpha}$ that is parallel propagated along its world line. It is straightforward to show that
\begin{equation}\label{W3}
e^{\mu}{}_{\hat \alpha} =  \delta^\mu_\alpha -\frac{1}{2} \,\bar{h}^{\mu}{}_{\alpha}\,.
\end{equation}
The projection of the curvature tensor~\eqref{G31} in the case of the incident gravitational wave  on the tetrad frame~\eqref{W3} results in $R_{\hat \alpha \hat \beta \hat \gamma \hat \delta} = R_{\mu \nu \rho \sigma}\,e^{\mu}{}_{\hat \alpha}\,e^{\nu}{}_{\hat \beta}\,e^{\rho}{}_{\hat \gamma}\,e^{\sigma}{}_{\hat \delta}$. Here, $e^{\mu}{}_{\hat \alpha}$ is in effect $\delta^\mu_\alpha$ in our linear approximation scheme and the GEM components of curvature can be represented as in Equation~\eqref{G28} with 
\begin{equation}\label{W4}
\mathbb{E}_{ij} = \frac{1}{2} \,\omega_g^2\, \bar{h}_{ij}\,, \qquad \mathbb{H}_{ij} = \frac{1}{2}\,\omega_g^2
\begin{bmatrix}
0 & 0 & 0 \\
0 & h_{\times} & -h_{+} \\
0 & -h_{+} &- h_{\times}
\end{bmatrix}
\,.
\end{equation}

For measurement purposes, it proves interesting to set up a quasi-inertial Fermi normal coordinate system with coordinates $X^{\hat \mu} = (cT, \widehat X,  \widehat Y,  \widehat Z)$ based on the nonrotating tetrad frame~\eqref{W3} along the world line of an arbitrary fiducial static geodesic  observer. Here, $T = t$ is the proper time of the reference observer fixed at $(x, y, z) = (x_0, y_0, z_0)$. The spacetime metric in the Fermi frame is given by
\begin{equation}\label{W5}
- ds^2 = g_{\hat \mu \hat \nu}\,dX^{\hat \mu} dX^{\hat \nu}\,
\end{equation}
where
\begin{equation}\label{W6}
g_{\hat 0 \hat 0} = - 1 - R_{\hat 0 \hat i \hat 0 \hat j}\,X^{\hat i} X^{\hat j}\,,
\end{equation}
\begin{equation}\label{W7}
g_{\hat 0 \hat i} =  - \frac{2}{3}\,R_{\hat 0 \hat j \hat i \hat k}\,X^{\hat j} X^{\hat k}\,
\end{equation}
and
\begin{equation}\label{W8}
g_{\hat i \hat j} = \delta_{ij} - \frac{1}{3}\,R_{\hat i \hat k \hat j \hat l}\,X^{\hat k} X^{\hat l}\,.
\end{equation}
In these expansions, we have neglected third and higher-order terms. In close analogy with the GEM case, we can define the gravitoelectric  potential $\hat{\Phi}$ and gravitomagnetic vector potential $\hat{\mathbf{A}}$ via $g_{\hat 0 \hat 0} = -1 + 2 \hat{\Phi}$ and $g_{\hat 0 \hat i} = - 2 \hat{A}_i$; that is, 
\begin{equation}\label{W9}
\hat{\Phi} = -\frac{1}{2}\, R_{\hat 0 \hat i \hat 0 \hat j}\,X^{\hat i} X^{\hat j}\,, \qquad \hat{A}_i =  \frac{1}{3}\,R_{\hat 0 \hat j \hat i \hat k}\,X^{\hat j} X^{\hat k}\,.
\end{equation}
Similarly, the corresponding fields can be defined as in Equation~\eqref{G4}; in fact, to lowest order we find
\begin{equation}\label{W10}
\hat{E}_i = R_{\hat 0 \hat i \hat 0 \hat j}\,X^{\hat j}\,, \qquad \hat{B}_i =  -\frac{1}{2}\,\epsilon_{ijk}\,R^{\hat j \hat k}{}_{\hat 0 \hat l}\,X^{\hat l}\,.
\end{equation}

Concentrating on the incident gravitational wave under consideration in this section, Equation~\eqref{W4} implies
\begin{equation}\label{W11}
\hat{\Phi} = -\frac{1}{4}\,\omega_g^2\,[h_{+} (\widehat{Y}^2-\widehat{Z}^2) + 2h_{\times} \widehat{Y}\widehat{Z}]\,
\end{equation}
and 
\begin{equation}\label{W12}
\hat{A}_1 = \frac{2}{3}\,\hat{\Phi}\,, \quad \hat{A}_2 = \frac{1}{6}\,\omega_g^2\, \widehat{X}\,(h_{+} \widehat{Y} + h_{\times} \widehat{Z})\,, \quad \hat{A}_3 = \frac{1}{6}\,\omega_g^2\, \widehat{X}\,(h_{\times} \widehat{Y} - h_{+} \widehat{Z})\,.
\end{equation}
Moreover, the relevant GEM fields are 
\begin{equation}\label{W13}
\hat{E}_1 = 0\,, \qquad \hat{E}_2 = \frac{1}{2}\,\omega_g^2\,(h_{+} \widehat{Y} + h_{\times} \widehat{Z})\,, \qquad \hat{E}_3 = \frac{1}{2}\,\omega_g^2\,(h_{\times} \widehat{Y} - h_{+} \widehat{Z})\,,
\end{equation}
\begin{equation}\label{W14}
\hat{B}_1 = 0\,, \qquad \hat{B}_2 = -  \hat{E}_3\,, \qquad \hat{B}_3 = \hat{E}_2\,,
\end{equation}
which are clearly transverse to the direction of wave propagation, $|\hat{\mathbf{E}}| =  |\hat{\mathbf{B}}|$ and $\hat{\mathbf{E}}\cdot \hat{\mathbf{B}} = 0$. For the incident wave, the gravitoelectric and gravitomagnetic potentials  are defined via the $g_{\hat 0 \hat \mu}$ components of the Fermi metric and the remaining spatial components can be expressed as
\begin{equation}\label{W15}
(g_{\hat i \hat j}) = 
\begin{bmatrix}
1+ \hat{A}_1 & \hat{A}_2 & \hat{A}_3 \\
\hat{A}_2  & 1-\xi_{+} & -\xi_{\times} \\
 \hat{A}_3 &  -\xi_{\times} & 1+\xi_{+}
\end{bmatrix}
\,,
\end{equation}
where 
\begin{equation}\label{W16}
\xi_{+} =  \frac{1}{6}\,\omega_g^2\,h_{+}\,\widehat{X}^2\,, \qquad \xi_{\times} =  \frac{1}{6}\,\omega_g^2\,h_{\times}\,\widehat{X}^2\,.
\end{equation}

Within this Fermi coordinate system, let us imagine the class of observers that stay spatially at rest. It is straightforward to show that a proper orthonormal tetrad frame $\Lambda^{\hat \mu}{}_{\tilde{\alpha}}$  adapted to this class of observers is given in $(t, \widehat X,  \widehat Y,  \widehat Z)$ coordinates by
\begin{equation}\label{W17} 
\Lambda^{\hat \mu}{}_{\tilde 0} = (1+\hat{\Phi}, 0, 0, 0)\,, 
\end{equation}
\begin{equation}\label{W18} 
\Lambda^{\hat \mu}{}_{\tilde 1} = (-2\hat{A}_1, 1-\tfrac{1}{3}\hat{\Phi}, 0, 0)\,, 
\end{equation}
\begin{equation}\label{W19} 
\Lambda^{\hat \mu}{}_{\tilde 2} = (-2\hat{A}_2, -\hat{A}_2, 1+\tfrac{1}{2}\xi_{+}, 0)\,, 
\end{equation}
\begin{equation}\label{W20} 
\Lambda^{\hat \mu}{}_{\tilde 3} = (-2\hat{A}_3, -\hat{A}_3, \xi_{\times}, 1-\tfrac{1}{2}\xi_{+})\,, 
\end{equation}
where the tetrad axes are primarily along the Fermi coordinate axes. 

Consider now a spinning test particle held fixed in space at $(\widehat X,  \widehat Y,  \widehat Z)$ by a reference observer in the Fermi frame. Projecting the spin vector $S^{\hat \mu}$ in the Fermi frame on the tetrad frame $\Lambda^{\hat \mu}{}_{\tilde \alpha}$ of the local reference observer, $S_{\tilde \alpha} = S_{\hat \mu} \Lambda^{\hat \mu}{}_{\tilde \alpha}$, we find $S_{\tilde 0} = 0$, as before, and 
\begin{equation}\label{W21} 
\frac{dS_{\tilde i}}{d\tilde{t}} = \left[\frac{D\Lambda^{\hat \mu}{}_{\tilde i}}{d\tilde{t}}\,\Lambda_{\hat\mu \tilde j}\right]\,S^{\tilde j}\,,
\end{equation}
where $\tilde{t}$ is the proper time of the reference observer and $dt = (1+\hat{\Phi}) d\tilde{t}$.  A detailed calculation reveals that to lowest order in $\widehat X$,  $\widehat Y$ and $\widehat Z$ within the Fermi coordinate system
\begin{equation}\label{W22} 
\frac{D\Lambda^{\hat \mu}{}_{\tilde i}}{d\tilde{t}}\,\Lambda_{\hat \mu \tilde j} = \partial_j\,\hat{A}_i -  \partial_i\,\hat{A}_j\,;
\end{equation}
hence, 
\begin{equation}\label{W23} 
\frac{dS_{\tilde i}}{d\tilde{t}} =  \epsilon_{ijk} \hat{B}^j\,S^{\tilde k}\,.
\end{equation}
Thus, as before, the dominant effect is that the spin vector precesses with an angular velocity given by the local gravitomagnetic field. We note that  $\Lambda^{\hat \mu}{}_{\tilde \alpha}$ differs from 
$\delta^\mu_\alpha$ by terms linear in the perturbation; hence, the gravitomagnetic field in Equation~\eqref{W23} is in effect the field measured by the reference observer. 
The corresponding Stern--Gerlach force, $f_{\hat \mu} = -\partial_{\hat \mu} (\mathbf{S}\cdot\hat{\mathbf{B}})$, to lowest order in $\widehat X$,  $\widehat Y$ and $\widehat Z$ as measured by the reference observer, is $f_{\tilde 0} = 0$ and
\begin{equation}\label{W24} 
f_{\tilde 1} = 0\,, \qquad f_{\tilde 2} = \tfrac{1}{2} \omega_g^2(h_{\times}S^{\tilde 2} - h_{+}S^{\tilde 3})\,, \qquad f_{\tilde 3} = -\tfrac{1}{2} \omega_g^2(h_{+}S^{\tilde 2} + h_{\times}S^{\tilde 3})\,.
\end{equation}
On the other hand, the Mathisson force~\eqref{G11} as measured by the reference observer is given by $F_{\tilde 0} = 0$ and
\begin{equation}\label{W25} 
F_{\tilde i} = \mathbb{H}_{\tilde i \tilde j}\,S^{\tilde j}\,,
\end{equation}
where $\mathbb{H}_{\tilde i \tilde j}$ is given to lowest order by Equation~\eqref{W4}. This is a consequence of the fact that in our approximation scheme $\Lambda^{\hat \mu}{}_{\tilde \alpha}$ is in effect given by $\delta^\mu_\alpha$ for the calculation of the measured components of the curvature tensor. It is then evident that the resulting components of the Mathisson force for the gravitational wave field under consideration in this section coincide with those of the Stern--Gerlach force given by Equation~\eqref{W24} in the correspondence limit.

\section{Discussion}

The Mathisson--Papapetrou equations for a spinning test particle together with the Frenkel--Pirani supplementary condition imply that the spin vector of a test pole-dipole particle is Fermi--Walker transported along its world line~\cite{Pirani}. For a spinning test particle held spatially at rest by a fiducial observer in the Ricci-flat region of an arbitrary gravitational field within the framework of linearized general relativity, the Fermi--Walker equation for the spin vector indicates  that its measured components undergo a precessional motion with an angular velocity that is given by the locally measured gravitomagnetic field. For an intrinsic quantum spin, there is therefore a spin-gravitomagnetic field coupling Hamiltonian associated with such precessional motion that can be obtained from Heisenberg's equation of motion. The gravitomagnetic field generally depends upon position; therefore, there is an accompanying Stern--Gerlach force connected with such a spin-gravity coupling. We show that under appropriate conditions, this Stern--Gerlach force reduces in the correspondence limit to Mathisson's classical spin-curvature force.

\section*{Acknowledgments}

I am grateful to Friedrich Hehl and Yuri Obukhov for valuable discussions.

\end{document}